\documentclass[12pt]{iopart}
\usepackage{bm}

\usepackage{graphicx} 
\usepackage{epstopdf}  
\usepackage{hyperref}
\usepackage{booktabs} 
\usepackage{dcolumn}
\usepackage{array} 
\usepackage{paralist} 
\usepackage{verbatim} 
\usepackage{color}

\usepackage{xcolor}

\usepackage{amssymb}

\usepackage[normalem]{ulem}

\newcommand{\eqref}{\ref}

\newcommand{\fig}{Fig.~}
\newcommand{\eq}{Eq.~}

\newcommand{\bu}{\mathbf{u}}
\newcommand{\bv}{\mathbf{v}}
\newcommand{\bx}{\mathbf{x}}
\newcommand{\bbr}{\mathbf{r}}
\newcommand{\bF}{\mathbf{F}}
\newcommand{\bxi}{\bm{\xi}}
\newcommand{\be}{\mathbf{e}}

\newcommand{\partfrac}[2]{\frac{\partial {#1}}{\partial {#2}} }

\newcommand{\ii}{{\rm i}}

\newcommand{\bzed}{\mathbf{0}}

\newcommand{\sep}{ \ \ \ , \ \ \ }

\newcommand{\bh}{{\bf h}}
\newcommand{\boldf}{\bf f}
\newcommand{\bV}{\bf V}

\newcommand{\beq}{\begin{equation}}
\newcommand{\eeq}{\end{equation}}
\newcommand{\beqn}{\begin{eqnarray}}
\newcommand{\eeqn}{\end{eqnarray}}
\newcommand{\pp}{\partial}
\newcommand{\dd}{{\rm d}}

\newcommand{\cO}{{\cal O}}

\newcommand{\denseVar}{\widetilde{\Delta \rho}}

\newcommand{\cutoffCILAlign}{0.024}
\newcommand{\cutoffCILDense}{0.012}
\newcommand{\cutoffCILZeta}{0.1}

\begin{document}

\title[Novel phase transitions in dense dry polar active fluids using a LBM]{Uncovering novel phase transitions in dense dry polar active fluids using a lattice
		Boltzmann method}

\author{David Nesbitt$^1$, Gunnar Pruessner$^2$, and
	Chiu Fan Lee$^1$}
\address{$^1$ Department of Bioengineering, Imperial College London, South Kensington Campus, London SW7 2AZ, U.K.}
\address{$^2$ Department of Mathematics,  Imperial College London, South Kensington Campus, London SW7 2AZ, U.K.}
 \ead{c.lee@imperial.ac.uk}

\date{\today}


\begin{abstract}
	The dynamics of dry active matter have implications for a  diverse collection of biological phenomena spanning a range of length and time scales, such as animal flocking, cell tissue dynamics, and swarming of inserts and bacteria. Uniting these systems are a common set of symmetries and conservation  laws,  defining dry active fluids as a class of physical system.
	Many interesting behaviours have been observed at high densities, which remain difficult to simulate due to the computational demand.
	Here, we show how two-dimensional dry active fluids in a dense regime can be studied using a simple modification of the lattice Boltzmann method. We apply our method on a model that exhibits motility-induced phase separation, and an active model with contact inhibition of locomotion, which has relevance to collective cell migration. For the latter, we uncover multiple novel phase transitions: two first-order and one potentially critical.
		We further support our simulation results with an analytical treatment of the hydrodynamic equations obtained via the Chapman-Enskog coarse-graining procedure.
\end{abstract}
\submitto{\NJP}
\maketitle



\section{Introduction} \label{sec:Introduction}

Symmetry serves a foundational role in all areas of physics today.
By identifying the underlying symmetries of a classical many-body system, one can derive the hydrodynamic equations of motion (EOM) that govern the  macroscopic dynamics of that system, and, crucially, any other system that respects the same symmetries  and conservation laws \cite{chaikin_lubensky1995}. In the case of simple (passive) fluids, temporal, rotational, translational, chiral and Galilean symmetries, and mass conservation lead to the Navier-Stokes equations in the hydrodynamic limit \cite{Forster1975}. Removing the Galilean symmetry and breaking the conservation of momentum lead instead to the Toner-Tu EOM \cite{Toner1995,Toner1998,Toner2012}, which generically describe dry polar active fluids  \cite{Marchetti2013}---systems typically composed of particles that generate non-momentum conserving propulsion through interaction with a fixed frictional medium.

Analysis of a hydrodynamic theory can elucidate the universal behaviour exhibited by all generic systems respecting the prescribed set of symmetries; conversely, any particular many-body system defined by microscopic rules that respect the same set of symmetries can also be used to study the associated universal behaviour in the hydrodynamic limit. An example of the former is the use of dynamic renormalisation group (DRG) methods to confirm the universal long-time tail phenomenon \cite{Forster1977} in thermal fluids first discovered by simulations \cite{Alder1967,Alder1970}; and an example of the latter is the use of lattice gas cellular automata to study the Navier-Stokes equations \cite{Frisch1986}. 
In dry active fluids, similar lattice gas models have helped to clarify the nature of the order-disorder transition in polar active fluids \cite{Csahok1995},  to elucidate the emergence of diverse dynamical structures \cite{Peruani2011},
and to determine the critical behaviour of motility-induced phase separation \cite{Partridge2019}. Superseding the lattice gas cellular automata  are the celebrated lattice Boltzmann methods (LBM) 
\cite{McNamara1988,Higuera1989,Higuera1989_enh,Chen1992,succi_rmp02}, which led to a drastic improvement in the computational efficiency of the simulation requirement.  

With their suitability to the almost incompressible regime, LBM have unsurprisingly been employed to simulate the fluid part of suspensions of  liquid crystals and active swimmers
\cite{Denniston2001,Denniston2004,Marenduzzo2005,Ramachandran2006,Marenduzzo2007,Cates2009,Doostmohammadi2015,Doostmohammadi2016,Doostmohammadi2017,Fielding2011,Giomi2008,Elgeti2011,Blow2014,Blow2017,Ravnik2013,Thampi2013,Thampi2014}. 		
 These examples all describe wet active matter with momentum conserving forces  \cite{Marchetti2013},
however the development of the LBM for dry polar active fluids is still lacking,  which is our focus here. 
Specifically, we present a method to study two-dimensional dry active fluids in a dense regime using a simple modification of the LBM. We then apply the simulation platform to a model that exhibits motility-induced phase separation (MIPS) and an active model with contact inhibition of locomotion (CIL). The latter is of direct relevance to migrating cells \cite{Theveneau2010,CarmonaFontaine2011}. We recover all known salient features of these two models. In addition, we uncover in the CIL model multiple novel phase transitions: two first order and one potentially critical. We further support our simulation results with an analytical treatment of the hydrodynamic equations obtained via the Chapman-Enskog coarse-graining procedure.

\begin{figure}
	\begin{center}
		\includegraphics[width=0.75\textwidth]{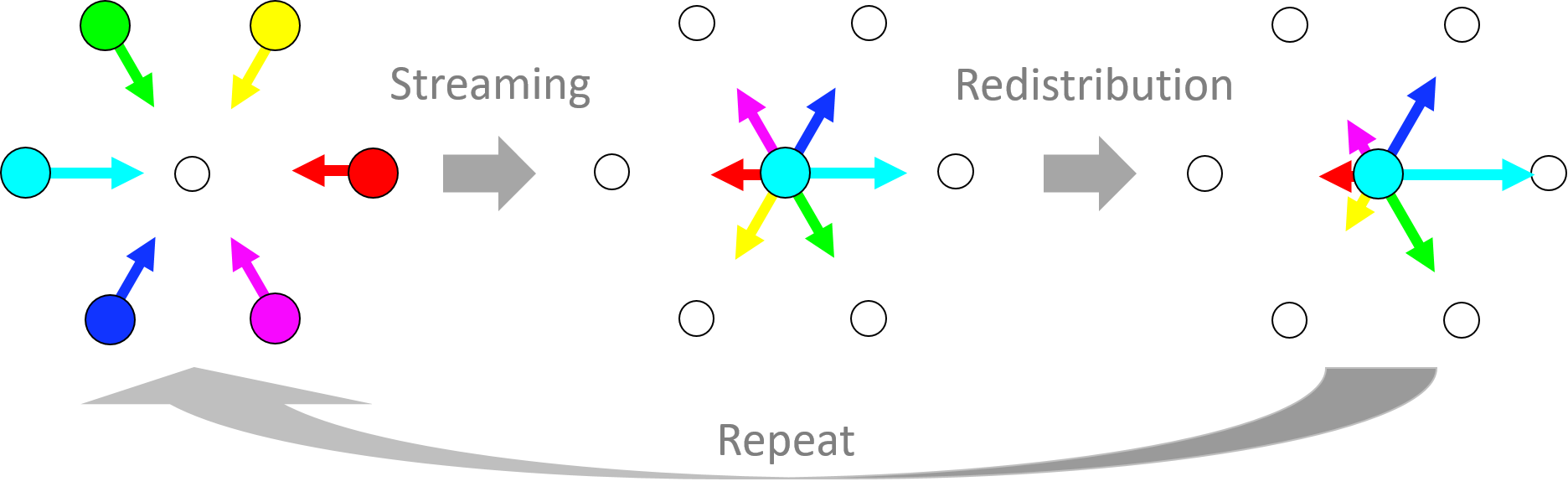}
	\end{center}
	\caption{
		Schematic of the temporal evolution of our lattice Boltzmann algorithm for a single site.
		 Streaming: Each directional density function propagates one lattice unit forward to the neighbouring site. Redistribution: At each lattice site, the resulting distribution $f_{i}$ is relaxed towards a steady state distribution $f_{i}^{\rm SS}$ that respects symmetries. 
	}
	\label{fig:LBD2Q7FullLBExplanationCompact}
\end{figure}

\section{Dry polar active fluids} \label{sec:DryPolarActiveFluids}
 Our system involves the evolution of a distribution function $f_{i}(t,\bbr)$,  $i \in \{0,1,...,6\}$, defined on a two-dimensional triangular lattice  (D2Q7 in standard LBM notation \cite{Qian1992})
  with corresponding lattice vectors $\be_{i}$ given by
${\be}_{0} = \bzed$, and  ${{\be}_{i}} = \cos [(i-1)\pi/6] \hat{\bx} +\sin [(i-1)\pi/6] \hat{{\bf y}}$ for $i>0$. 
The distribution function $f_i$ is related to the local mass density and local velocity  of the system as follows:
\begin{equation}
 	 \rho(t,\bbr) = \sum_{i=0}^{6} f_{i}(t,\bbr) 
 	 \sep
 	  \bu(t,\bbr) = \frac{\sum_{i=0}^{6} f_{i}(t,\bbr) c \be_{i}}{\rho(t,\bbr)}
 	  \ , \label{eqn:D2Q7macroscopic_u_disc}
\end{equation}  
where
 $c$ is the ratio of the lattice spacing $\Delta x $ to the time increment $\Delta t$. Typically, scales are chosen so that $c=1$, and $|\bu|\ll c$ is required for stable solutions (see \ref{sec:LBM}). 

We then evolve the discretised Boltzmann equation 
using the Bhatnagar-Gross-Krook collision operator \cite{Bhatnagar1954}, albeit with fluctuations included, as follows:
\begin{equation} \label{eqn:fevolution2}
	f_{i}( t + \Delta t , \bbr + c \be_{i} \Delta t ) - f_{i}(t,\bbr )  
	= 
	- \frac{1}{\tau} \left[ f_{i}(t,\bbr )   -  f_{i}^{\rm SS}(t,\bbr ) \right]  
	+ {\eta}_{i}(t,\bbr) \ ,
\end{equation}
where $\tau$ is a relaxation parameter, $f_{i}^{\rm SS}$ is the steady-state distribution, and
\begin{equation}
	\eta_{i}(t,\bbr) = \tilde{\eta}_{i}(t,\bbr) - \frac{1}{7}\sum_{i=0}^{6}\eta_{i}(t,\bbr) \label{eqn:noiseDensityConservation}
	\ ,
\end{equation}
where $\tilde{\eta}_{i}$  are temporally and spatially uncorrelated random variables uniformly distributed in the interval $[-\sigma, \sigma]$.
This particular choice of noise alters the local velocity, but \eq \eqref{eqn:noiseDensityConservation} ensures the density is conserved. 
Note that the fluctuations do not conserve momentum as we are considering dry active fluids.

In standard LBM, $f_{i}^{\rm SS}$ is constructed so as to preserve the moments $\rho$, $\bu$ at each site, accomplished in a D2Q7 model by
\begin{equation}
	f_{i}^{\rm SS} = w_i \rho \left[ 1	 +  4\frac{\be_i \cdot \bu^{*} }{c} + 8\frac{(\be_i \cdot \bu^{*} )^2}{c^2}- 2\frac{|\bu^{*} |^2}{c^2}	  \right] 
	\ , \label{eqn:fssD2Q7}
\end{equation}
with $\bu^{*} =\bu$, where $w_i$ are lattice specific weights: $w_0 = 1/2$ and $w_{i\ne0} = 1/12$.
However,  in breaking the conservation of local momentum, the restriction on the form of the steady-state distribution $f_{i}^{\rm SS}$ is  drastically relaxed.
 We  make a simple modification:  maintaining the form of $f_{i}^{\rm SS}$ in \eq \eqref{eqn:fssD2Q7}, but taking $\bu^{*}$ to be a function  of  $\rho$ and  $\bu$,  that respects the symmetries of our system of interest.
This update rule conserves mass density, but does not conserve momentum nor energy, thus rendering the system in the realm of the hydrodynamic theory of Toner and Tu, rather than Navier-Stokes  (see \ref{sec:TonerTuRecovery}).
 This method can produce a variety of behaviours associated with dry active matter, depending on the choice of $\bu^{*}$.

\section{Motility-Induced Phase Separation} \label{sec:MIPS}


To demonstrate the versatility of our method, we first employ it to produce motility-induced phase separation (MIPS). MIPS can occur when self-driven particles that maintain their direction of propulsion with some persistency, interact through short-range repulsive forces \cite{Cates2015,Fily2012,Bialke2013,Buttinoni2013,Redner2013,Stenhammar2013,Stenhammar2014,Partridge2019}. The particles congregate in high density clusters, resembling a liquid-gas phase separation.

The phase separation arises from two related mechanisms. Firstly, it has been observed that active particles tend to congregate in regions where their speeds are slow \cite{Schnitzer1993}. This is well demonstrated using light sensitive strains of {\it E. coli} bacteria, with heterogeneous distributions of light leading to matching density profiles \cite{Arlt2018}.
Secondly, in systems with repulsive forces, the converse is also true; in areas with high densities, the speeds slow down. 
The exact reason for the slowdown differs depending on the system of interest. Run-and-tumble bacteria respond to chemical secretions by their neighbours, which affects their own behaviour \cite{Miller2001,Meyer2014}. Active Brownian particles on the other hand do not alter their behaviour in the presence of other particles. Rather, at high densities, they experience many collisions in a short time, disrupting their motion. If their trajectory is averaged over a short interval, they are seen to effectively slow down at higher densities \cite{Fily2012,Bialke2013}.

These two mechanisms combine to form a positive feedback loop, and an instability occurs. Particles in a region of high density slow down, resulting in more particles congregating in that region, increasing the density further. The cluster grows and merges with others until eventually a stable equilibrium in the system is achieved.


To generate MIPS using our method we take
\begin{equation} \label{eqn:bu_mips} 
	\bu^{*} =  
	\left\{
	\begin{array}{cc}
		C \left(1 - \frac{\rho}{\rho^{*}} \right)\nabla \rho \ , & {\rm if } \rho < \rho^{*},\\
		0 \ , & {\rm otherwise},
	\end{array}
	\right . 
\end{equation}
where $C$ is a positive constant, dictating the sensitivity of the system to density gradients. The parameter $\rho^{*}$ acts as an upper bound for the local density as permitted by the system; above this there is no net flow, regardless of density gradients, hence the density cannot rise further.
Since $\nabla \rho$ could in principle become very large and any physical system will not have a divergent speed, we impose further an upper bound on $|\bu^{*}|$. Explicitly, we have 
\begin{equation}
	|\bu^{*}| \mapsto \min\left( |\bu^{*}| , \ U_{\rm max} \right) \ .
\end{equation}

Standard LBM do not typically feature gradients. Indeed, apart from the streaming step, all of the calculations and operations are strictly localised to each lattice site. To calculate a gradient we must include information from neighbouring nodes. We do so by incorporating a finite difference procedure into the routine \cite{Swift1996}.
It is important to use a finite difference stencil that is isotropic when projected to the lattice. 
A suitable choice is given by
\begin{equation}
	\nabla \rho = a \left[  (\rho_{1} - \rho_{4})\be_{1}  +  (\rho_{2} - \rho_{5})\be_{2} +  (\rho_{3} - \rho_{6})\be_{3}     \right]  \ ,
\end{equation}
for $a=1/(3\Delta x)$, where in this notation $\rho_{i} = \rho(t, \bbr + \Delta x \be_{i})$ \cite{Swift1996}.
The resulting stencils are
\begin{eqnarray}
	\partfrac{\rho_0}{x} &= \frac{2(\rho_1 - \rho_4) + (\rho_2 + \rho_6) - (\rho_3 + \rho_5) }{6 \Delta x} , \\ \partfrac{\rho_0}{y} &= \frac{(\rho_2 + \rho_3) - (\rho_5 + \rho_6) }{4\Delta y} \ .
\end{eqnarray}

We motivate the expression in \eq \eqref{eqn:bu_mips} firstly by intuition, and secondly through comparison with other studies.
For our system to experience MIPS the speed of particles must decrease in regions of high density. The speeds of individual particles do not feature in our system, as $\bu^{*}$ is a macroscopic variable. However, a consequence of all particles' speeds decreasing is that the net speed $|\bu^{*}|$ will also decrease.
This is clearly achieved by the term $\left(1 - {\rho}/{\rho^{*}} \right)$, with the net speed hitting zero altogether for $\rho\ge \rho^{*}$. Physical systems of hard particles have an upper limit as to how tightly they can be packed, hence it is natural to have a maximum density $\rho^{*}$.
There is no alignment mechanism in a MIPS model and directions of particles are typically random and isotropic. This raises a question as to how we can have a non-zero net flow at all.
The answer comes in the form of a statistical average. 
Suppose at some location in our system there is a non-zero density gradient. Nearby particles travelling in the same direction as the gradient will be entering a region of higher density, therefore they will experience more collisions and thus be impeded more than particles moving in other directions. As the particles maintain their current direction with some persistency, after a short time the impeded particles will still be in place, pointing towards the density gradient, whilst the non-impeded particles will have left the vicinity entirely. The average velocity of the particles will therefore align with the density gradient, and as a result there is a net flow towards regions of higher density.
This is also consistent with the observation that particles will congregate in areas of lower speeds or, equivalently, higher densities; we expect there to be a flow towards those regions.

We can also arrive at the expression in \eq \eqref{eqn:bu_mips} through comparison with previous studies in the field. In the notation of \cite{Cates2015}:
for a system of non-interacting active particles propagating with fixed speed $v(\bbr)$, and orientational relaxation time $\tau(\bbr)$, where $v$ and $\tau$ can vary in space, it has been shown that the macroscopic density can be approximated by the following
	\begin{eqnarray}
		\partfrac{\rho}{t} &= - \nabla \cdot \mathbf{J} \ , \label{eqn:MIPS_Macro_Continuity} \\
		\mathbf{J} &= - D \nabla \rho + \bV \rho + \sqrt{2D\rho}\Lambda \ ,  \label{eqn:MIPS_Macro_J}
	\end{eqnarray}
where $\bV$ is a drift velocity, $D$ is a diffusivity, and $\Lambda$ is a white noise \cite{Cates2015}. With negligible translational diffusivity these can be expressed as
\begin{eqnarray}
	D &= \frac{v^2\tau}{d} \ , \\
	\frac{\bV}{D} &= - \nabla \log v \ ,
\end{eqnarray}
where $d$ is the dimension.
Generalizing this system to include interaction effects, allows $v$ and $\tau$ to now depend on the local density $\rho$, as well as $\bbr$. By extension so do $D$ and $\bV$.
For simplicity we assume $\tau$ to be constant and $v$ to depend only on $\rho$. Numerical studies of active Brownian particles have shown that the effective speed has an approximately linear dependence on the density \cite{Fily2012,Bialke2013,Stenhammar2013,Stenhammar2014}. 
Hence we adopt
\begin{equation}
	v(\rho) = v_0 \left(1 - \frac{\rho}{\rho^{*}} \right) \ .
\end{equation}
The drift velocity $\bV$ becomes
\begin{equation}
	{\bf V} = \frac{\tau v_0^2}{\rho^{*} d} \left(1 - \frac{\rho}{\rho^{*}} \right)\nabla \rho \ .
\end{equation}

Finally we observe that \eq \eqref{eqn:MIPS_Macro_Continuity} is a continuity equation and equate $\mathbf{J} \equiv \rho \bu$ in our system. We take the drift velocity to be the steady state that $\bu$ tends towards, that is $\bV \equiv \bu^{*}$. The term $D \nabla \rho$ features in our model as the ``pressure'' term in our hydrodynamic equations (see \ref{sec:TonerTuRecovery}), and we further incorporate fluctuations in the system. Our system is very similar to that presented in \eq \eqref{eqn:MIPS_Macro_Continuity}, \eqref{eqn:MIPS_Macro_J}, albeit our velocity $\bu$ is dynamic, whilst $\mathbf{J}$ has no time dependence. In that sense, \eq  \eqref{eqn:MIPS_Macro_J} is the overdamped version of our model.

\begin{figure*}
	\begin{center}
		\includegraphics[width=\textwidth]{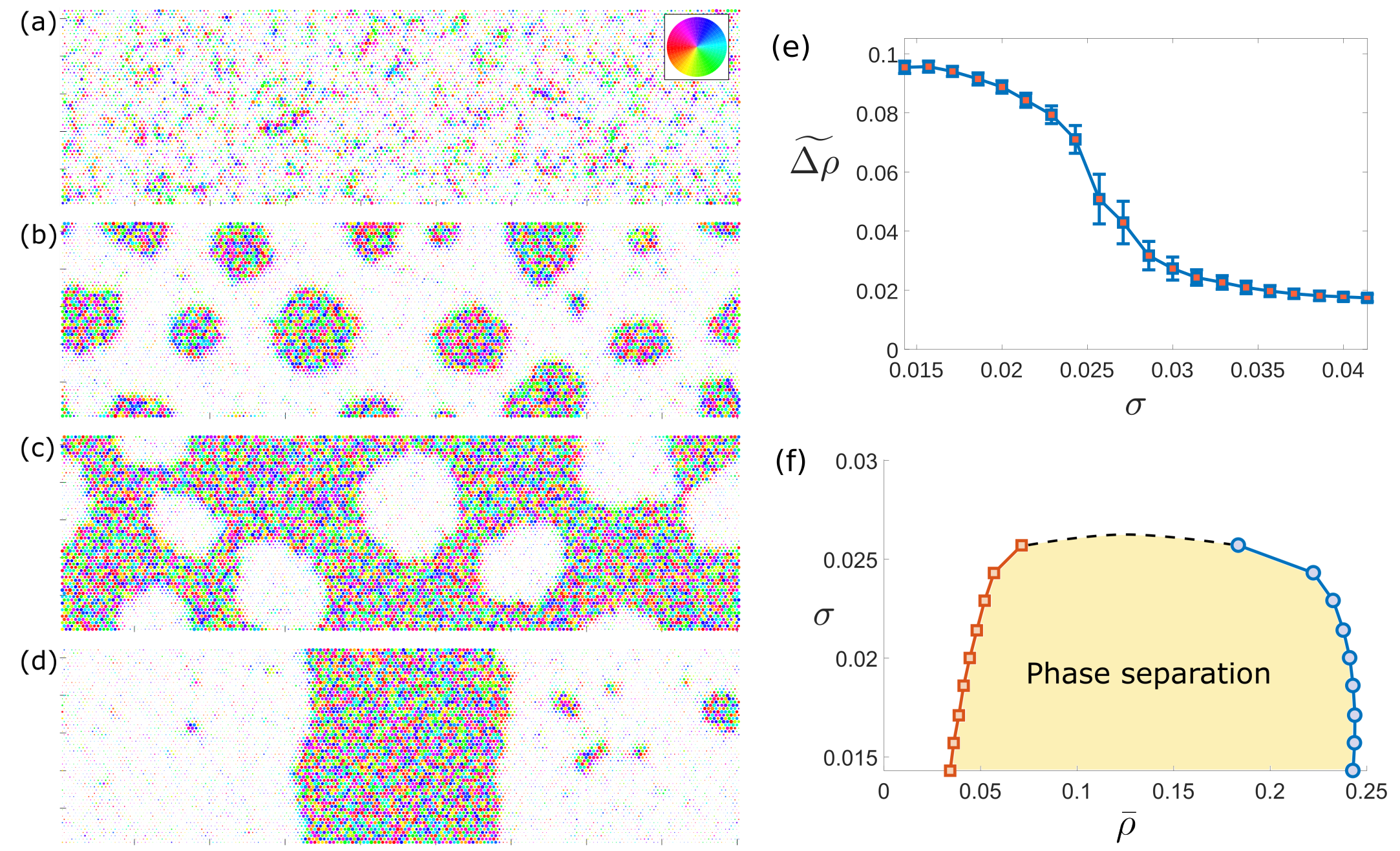}
	\end{center}
	\caption{
		Results obtained using our LBM for MIPS. See Table \ref{tab:ParametersMIPS} for parameter values shared by all subfigures. (a)-(d) Example behaviour of the system; each image depicts the state of the entire lattice for a single time step. The plots show a coloured dot for each of our $180\times60$ triangular lattice sites, with the size of the dot depicting the density $\rho$ and the colour denoting the direction of $\bu$ according to the inset colour wheel. In (a) the noise is too strong for stable phase separation to occur. Frames (b)-(d) demonstrate MIPS. (b) shows liquid clusters distributed in a gaseous domain. (c) has a higher mean density than (b), resulting in gas bubbles within a liquid domain. (d) is from the same simulation as (b), but shows a much later time step, at which point the system has almost reached the theoretical equilibrium.
		(e) shows the dependence of $\denseVar$ on the noise strength. We use $\denseVar$ to indicate whether or not phase separation has occurred (see \ref{sec:RegimeCategorisation}).
		(f) is a phase diagram. For $(\bar{\rho},\sigma)$ pairs that fall within the yellow region, phase separation occurs. For $\sigma > \sigma_{c}\approx 0.026$ no phase separation can possibly occur.
	}
	\label{fig:MIPS}
\end{figure*}


We ran our simulation in a rectangular geometry with periodic boundary conditions. For simulation details, including parameter values, see \ref{sec:SimulationDetails}.
\fig \ref{fig:MIPS} shows the results from our MIPS simulations. \fig \ref{fig:MIPS}(a)-(d) are still frames from our simulations displayed using the visualization scheme described in \ref{sec:VisualizationScheme}. In panels (b)-(d) MIPS has occurred. The multi-coloured nature of the the dense region indicates there is no velocity alignment between neighbouring lattice sites. Therefore, collectively, the dense regions have effectively zero velocity. 

\fig \ref{fig:MIPS}(f) displays a phase diagram, where the orange squares are the mean gas densities, and the blue circles are the mean liquid densities measured in systems that exhibited phase separation. Error bars for the standard error of the mean are smaller than the marker size.
The data points appear to trace out a convex hull, the uppermost limits of which we have estimated with the black dashed line.
Any system whose mean density and noise strength, $(\bar{\rho},\sigma)$, fall within the yellow region will experience phase separation, with the density in their liquid and gas phases given by the values on the blue and orange curves for that $\sigma$, respectively.

These results are consistent with previous studies in MIPS \cite{Partridge2019}. As such, we have demonstrated that our lattice Boltzmann method is adaptable, and could be used to study MIPS in more detail.


\section{Contact Inhibition of Locomotion} \label{sec:CIL}

The remainder of this paper is concerned with a class of active systems that exhibit collective motion, for which we take 
\begin{equation} \label{eqn:bu_CIL}
	\bu^{*} = \frac{U_{0}}{|\bu|}\bu \ , \\
\end{equation}
where $U_0$ is of the form
\begin{equation}
	\label{eqn:U0}
	U_0(\rho) =
	\left\{
	\begin{array}{cc}
		\hfill -A\rho^2+B & {\rm if } B \ge A\rho^2 \ ,\\
		\hfill 0 & {\rm otherwise}.
	\end{array}  
	\right . 
\end{equation}
In the above,  $A,B$ are positive constants. We expect this to reproduce the dynamics of active polar matter because it mimics the key behaviour that local particles' movements tend to align.
As $f_{i}^{\rm SS}$ is defined at each lattice site, the alignment interactions are short-ranged.
In addition, we assume here that collective motility can be impeded when the density gets too high or even stop altogether,
which can be viewed as a form of contact inhibition of locomotion (CIL), as observed in cell monolayers \cite{Petitjean2010,Angelini2011,Garcia2015} and traffic jams \cite{Lighthill1955,Schnyder2017}.
By using a quadratic expression, we assume that contact inhibition effects are minimal for low densities, and only become relevant at higher values. We will show in Section \ref{sec:Renormalization} that incorporating fluctuations will generally render the effective ``speed'' function $U_0$ a much more complicated function of $\rho$.

\begin{figure*}
	\centering
	\includegraphics[width=\textwidth]{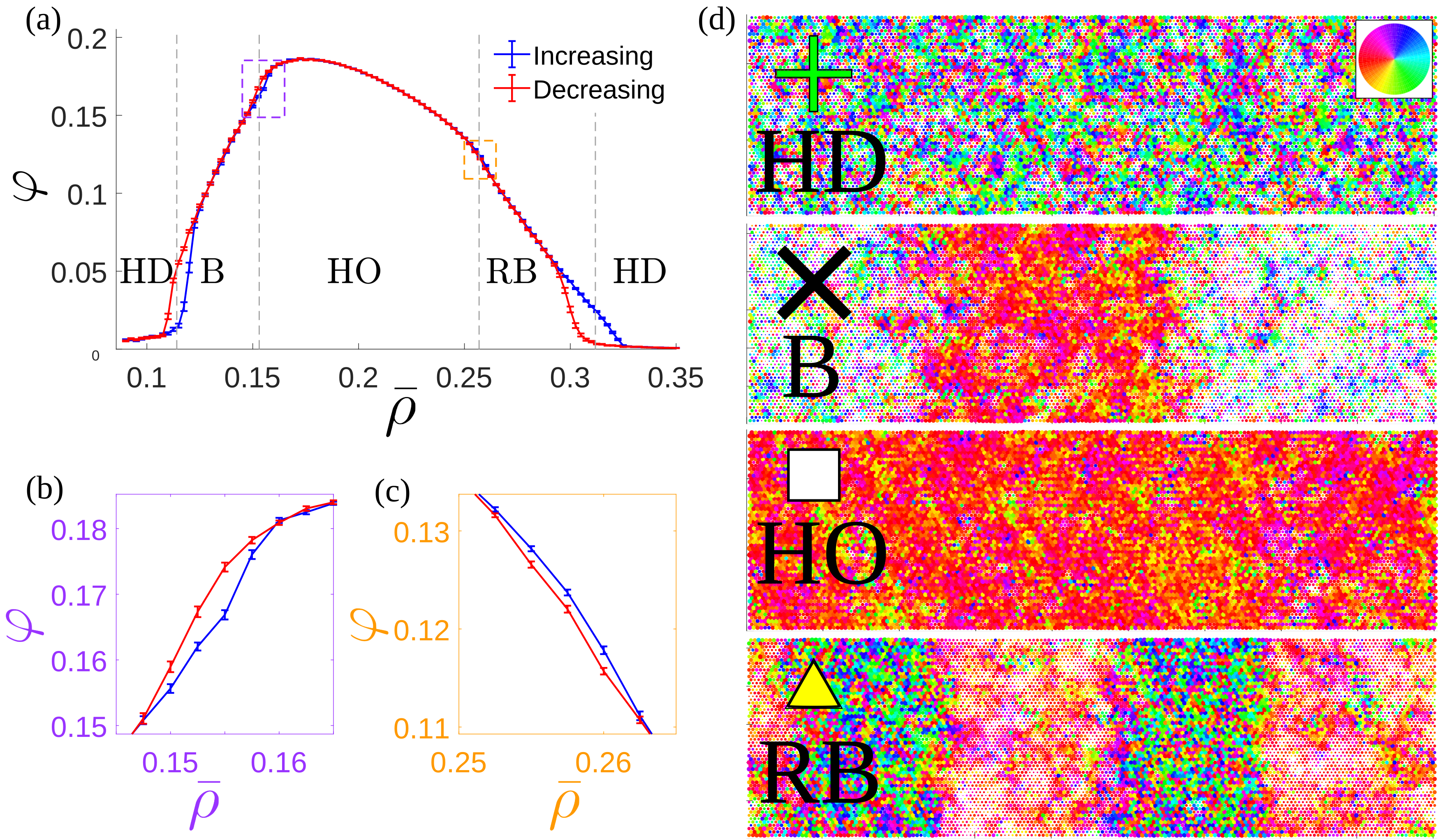}
	\caption{
		(a) Order parameter $\varphi$ (\ref{eq:phi})  vs.~the average density $\bar{\rho}$ for $A = 2, B=0.3$ (\ref{eqn:U0}). 
		Measurements were taken whilst slowly increasing (blue) or decreasing (red) the average density in the system. The results align with the exception of four hysteresis loops, which   locate  the corresponding first-order phase transitions. 
		The phase transitions partition the system into four distinct regimes: 
		HD -- homogeneous and disordered, B -- banding, HO -- homogeneous and ordered, RB -- reverse banding. 
		(b),(c) Magnified versions of the purple and orange dashed boxes respectively highlighting the hysteresis effect in each.
		(d) Snapshots of the typical steady state behaviour of each of the phases. Movies are available online (see \ref{sec:Movies}). The plots show a dot for each of our $180\times60$ triangular lattice sites, with  the size of the dot depicting the density $\rho$ and the colour denoting the direction of $\bu$ according to the inset colour wheel. White space implies a region of low density. 
		Coloured symbols are identifiers for the data points in \fig \ref{fig:PhaseDiagram}(a). See \ref{sec:SimulationDetails} for details on simulation procedure and parameter values.
	}
	\label{fig:Hysteresis}
\end{figure*}

As for our MIPS model, we simulate our active fluid model in a rectangular geometry with periodic boundary conditions (\fig \ref{fig:Hysteresis}(d)). This ensures that any ensuing collective motion aligns itself along the $x$-axis,  which helps to reduce noise in our data collection.
We use the average speed of the whole system as an order parameter $\varphi$, given by
\begin{equation}
	\label{eq:phi}
	\varphi
	= \left| \frac{\sum_{ \bbr \in  {\rm lattice}}\rho { (\bbr) \bu (\bbr)} }{\sum _{ \bbr \in  {\rm lattice}}\rho { (\bbr) }} \right|
	\ .
\end{equation}
Necessarily $0 \le \varphi \le B$. 
 \fig \ref{fig:Hysteresis}(a) shows how $\varphi$
 behaves with varying average density $\bar{\rho}$. Two curves have been traced by slowly increasing (blue) and decreasing (red) $\bar{\rho}$. The two curves coincide with the exception of four regions, in which there exist hysteresis loops indicating the existence of meta-stable states. We  will use these hysteresis loops as proxies for the locations of first-order  phase transitions that separate distinct phases.
 At low densities noise dominates, and the system is in the homogeneous and disordered (HD) phase. As the density increases, alignment effects take over and dense locally aligned bands spontaneously develop, which propagate through a low density  disordered region. 
 	We refer to this as the banding (B) regime, which corresponds to the coexistence of the HD phase in the dilute regions 
 and a homogeneous, ordered (HO) phase in the dense regions. 
 	 Increasing the density further causes the dilute HD phase to vanish,  resulting in a single HO phase.  
 	 The first HD-B transition is easily recognised from previous flocking studies \cite{Csahok1995,Chate2008,Gregoire2004,Bertin2006,Bertin2009,Thuroff2014} which found it to be first order.
The second B-HO transition is less studied, likely because higher density molecular dynamics simulations are more difficult to compute, but has recently been shown to also be a first order transition  \cite{Thuroff2014}.
Our results are in perfect agreement with both of these as evidenced by the aforementioned hysteresis loops.
This lends credence to the conclusion that the two remaining hysteresis loops at higher densities also indicate first-order phase transitions, in agreement with a recent experimental study concurrent to ours \cite{Geyer2019}.
These transitions separate the HO region from another HD region, again through a coexistence regime similar to the B regime (\fig \ref{fig:Hysteresis}(d)), but with the properties inverted---the high density regions are disordered, whilst the low density regions are aligned and moving.
In the steady state, as material flows into a dense band it comes to a halt, extending the band on that side, whilst on the other side the band decays as the material flows away.
The net result is that the dense band, although locally stationary, appears to propagate against the direction of flow, hence we call it reverse banding (RB).  In other words, the dense band moves due to condensation on one side and evaporation on the other. In active matter, this phenomenon was first reported in a collective cell migration model \cite{Schnyder2017}, and likened to density waves observed in traffic flow \cite{Lighthill1955}. Interestingly, the same mechanism also underlies the directed motion of condensed drops under the influence of a chemical gradient \cite{Weber2017}, as well as the formation of a drop lattice in a chemical reaction-controlled phase separated drop system \cite{LeeWurtz2019}.

 The RB regime also bears striking resemblance to the glider structure discovered in \cite{Peruani2011}, which also moves backward due to a similar differential absorption and evaporation mechanism. However, we believe that the RB regime discussed is fundamentally distinct from the glider for the following two reasons: i)  the glider  in \cite{Peruani2011} does not form a band but is triangular in shape; and ii) a glider consists of two halves, each with a distinct mean orientation. That is, the mean orientation differs within the glider depending on which region one focuses on, while the reverse band discussed here is ordered uniquely and homogeneously within the band.

\begin{figure}[h!]
	\centering
	\includegraphics[width=\textwidth]{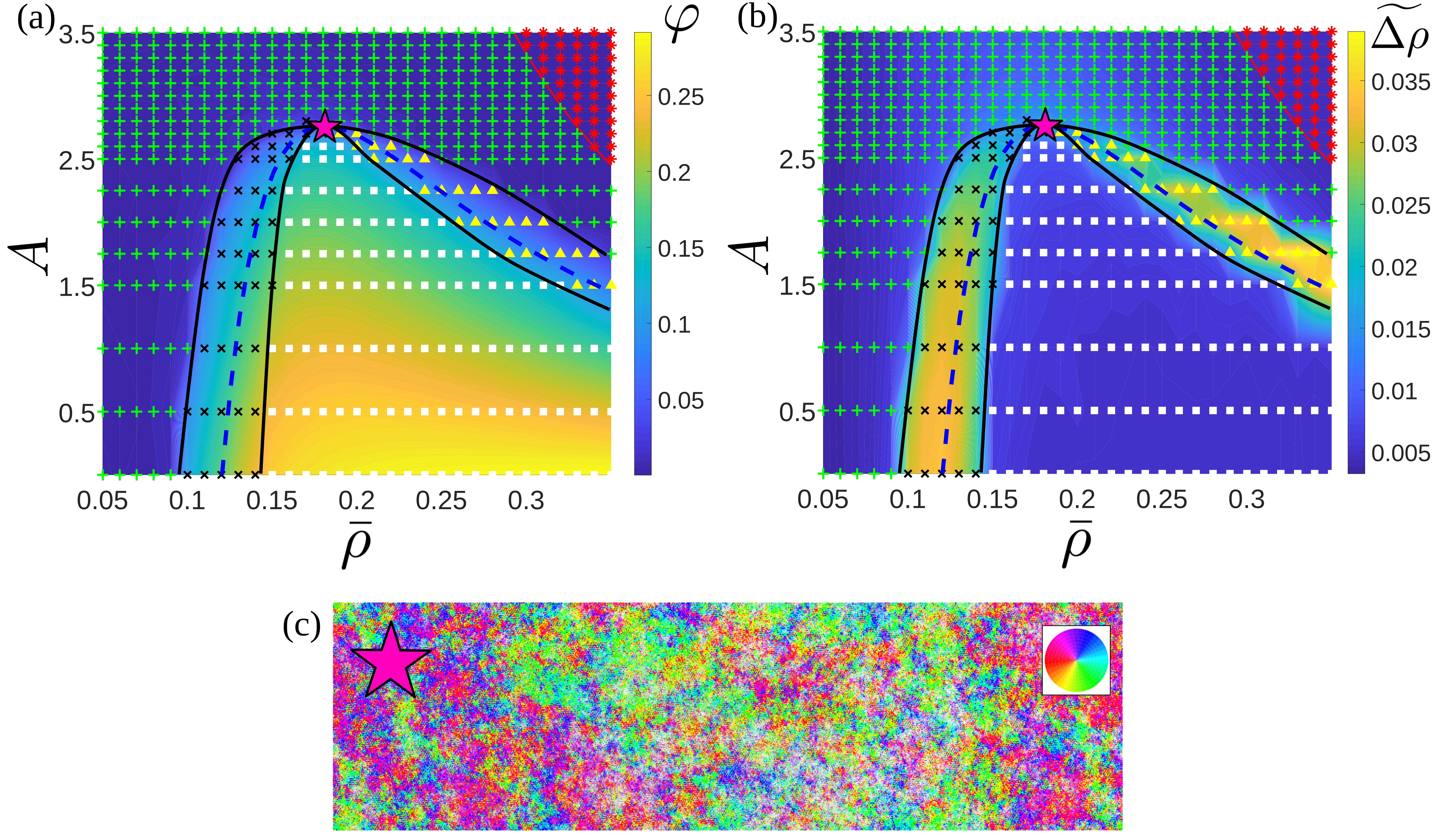}
	\caption{ 
		(a) Phase diagram comparing average density  $\bar{\rho}$ with coefficient $A$ from \eq \eqref{eqn:U0}. Each sample point was identified as belonging to one of five regimes, categorized by the coloured symbols (see \fig \ref{fig:Hysteresis}(d)) using a method detailed in \ref{sec:RegimeCategorisation}.
		The red asterisk region has zero motion because $|\bu| = B -A  \bar{\rho}^2 \le 0$ with equality depicted by the red line. 
		Black lines depict approximate locations of phase transitions (drawn by eye) which meet at a critical point (star).  Blue dashed line is the fluctuation-renormalized curve $\alpha_{R}=0$ (Section \ref{sec:Renormalization}) fitted to pass through the critical point. The banding (black cross) and reverse banding (yellow triangle) regions straddle this curve, as predicted by our stability analysis.
		  (b) Snapshot of behaviour close to the  critical point at  $ \bar{\rho}_c \approx 0.18$ and $A_c \approx 2.75$, for a lattice of size $720\times240$ and using the same display scheme as in \fig \ref{fig:Hysteresis}(d).
		  Simulation details are in \ref{sec:SimulationDetails}.	}
	\label{fig:PhaseDiagram}
\end{figure}

\section{Phase behaviour} \label{sec:PhaseBehavior}
To further elucidate the phase behaviour of the system, we now vary the contact inhibition parameter $A$ as well as $\bar{\rho}$ (\fig~\ref{fig:PhaseDiagram}(a)). 
When $A=0$,  we recover the expected phenomenology of a typical polar active model, such as the Vicsek model \cite{Vicsek1995}, in that the system  transitions from the HD phase to the B regime followed by the HO phase as density increases.
However, for $A>0$, we see that the HO phase will transition back to the HD phase via the RB regime as density increases further. 
Most interestingly, as $A$ increases, all four first-order phase transition lines converge to a single point, indicated by a star in \fig~\ref{fig:PhaseDiagram}(a), which is a putative critical transition.  
  A snapshot of this critical behaviour is shown in \fig \ref{fig:PhaseDiagram}(b).  We will discuss this critical point further
 in our hydrodynamic analysis.
Finally, we note that in addition to the four regimes already discussed we recognise a fifth somewhat trivial phase of no motion (indicated by red asterisks), where $|\bu| = 0$ since $- A \bar{\rho}^2 +B \le 0$.

We note that although we focus on the phase separations within the ordered phase, the interesting interplays between MIPS and the polar phase have also been discussed in various recent works \cite{farrell_prl12,Barre2015,MartinGomez2018,SeseSansa2018,VanderLinden2019,Caprini2020,Jayaram2020,Bertrand2020}, which are complimentary to this study.

\section{Hydrodynamic description of the transitions} \label{sec:HydroAnalysis}
In the hydrodynamic limit, our model is necessarily described by the Toner-Tu  equations due to our imposed symmetries, and we confirm this analytically using the Chapman-Enskog expansion method in \ref{sec:TonerTuRecovery}. We now 
describe our simulation findings analytically by performing a linear stability analysis on  our derived version of the Toner-Tu equations:
	\begin{eqnarray}
		\partial_{t} \rho  + \nabla \cdot \left( \rho \bv \right) = 0 \ ,	\\
		\fl \qquad	\rho \left(	\partial_{t} \bv   +  \bv \cdot \nabla   \bv \right)= - \nabla P   + \nu \nabla \cdot \rho  \left( \nabla \bv + (\nabla \bv)^{T} \right) 	
			+  \rho \left[\alpha(\rho)- \beta |\bv|^2 \right]\bv \ ,
	\end{eqnarray}
	where $P = c_{s}^2 \rho$.

	Since the bands in both the B and RB regimes are perpendicular to the direction of collective motion in the dense and dilute regions, respectively, 
	the most unstable wave vectors in the HO state are expected to be parallel to the direction of motion. 
	Therefore, we need only consider a one-dimensional system, where the coordinate $x$ is the direction of the collective motion. Expanding about a homogeneous collective motion state with 
	$\rho = \rho_0 + \delta \rho \exp (s t -  \ii kx )$, 
	$v =  \sqrt{\alpha_0/\beta} + \delta v \exp ( s t -  \ii kx )$ 
	and $\alpha  = \alpha_0 +\alpha_1 \delta \rho$, we have to linear order:
	\begin{eqnarray}
		&\qquad \qquad \left(s - \ii k \sqrt{\frac{\alpha_0}{\beta}} \right) \delta \rho = \ii k \rho_{0} \delta v \ , \\
			&\rho_{0} \left(s - \ii k \sqrt{\frac{\alpha_0}{\beta}} + 2 \alpha_0 + 2 \nu k^{2} \right) \delta v  
			 = \left(\ii k c_{s}^{2} + \rho_{0} \alpha_1 \sqrt{\frac{\alpha_0}{\beta}}\right) \delta \rho \ .
	\end{eqnarray}
	Note that for a stable system $\beta>0$ necessarily, thus $\alpha_0>0$ implies collective motion, with $\alpha_0 \le 0$ giving a stationary state.
	
	Solving for $s$ and focusing on the real part, which determines the stability of the system, we have
	\begin{equation} \label{eqn:BandingInstabilityCondition}
		{\rm Re}[s] 
		=\left\{
		\begin{array}{l}
		-2 \alpha_0 +\cO(k^2) \ ,
		\\
		\frac{k^2}{8 \alpha_0^2 \beta} 
		\left[ \alpha_1^2 \rho_{0}^{2}  - 4 \alpha_0 c_{s}^{2} \beta 
		\right]  +\cO(k^3) \ ,
		\end{array}
		\right.
	\end{equation}
	where ${\rm Re}[s]>0$ implies instability. 
	The first  eigenvalue $-2\alpha_0$ corresponds to the fast relaxation that occurs when the velocity deviates from the mean field value $\sqrt{\alpha_0/\beta}$ in the absence of spatial variations.
	The second eigenvalue quantifies when the instability sets in. In particular, if $\alpha_0 \to 0$ and $\alpha_1 \ne 0$, the system is always unstable. 
	In the previously mentioned studies, \cite{Csahok1995,Vicsek1995,Gregoire2004,Bertin2006,Bertin2009,Thuroff2014}, $\alpha_1$ is typically assumed to be positive, so that alignment interactions increase with density, and as $\alpha$ goes from negative to positive the system will necessarily become unstable, resulting in the banding instability. 
 	This corresponds to the banding (B) regime in our system, since a higher density reduces the effects of the noise, and as such alignment increases with density.
However, it is also clear that the instability condition does not depend on the sign of $\alpha_1$.
Hence, if $\alpha$ goes from positive to negative due to contact inhibition, the system will again be unstable.
This is exactly what occurs in our simulation and corresponds to the Reverse-Banding (RB) regime in our phase diagram. 
In our simulation we found that all four first-order phase transition lines converge at a critical point.
With our hydrodynamic argument, we can see that this critical point is  reached by setting both $\alpha_0$ and $\alpha_1$ to zero, which is achieved by fine-tuning $\bar{\rho}$ and $A$ in our simulation. 
 Furthermore, as detailed in the following section, by capturing the effect of fluctuations at the one-loop level, we can obtain an expression for $\alpha(\rho)$ in our system. 
 The blue dashed curve in \fig \ref{fig:PhaseDiagram}(a) was obtained by fitting the curve $\alpha_{R}(\bar{\rho})=0$ to pass through the critical point. The resulting curve bisects both the B and RB regions, which is in perfect agreement with our stability analysis; close to this line $\alpha_0 \approx 0$ and $\alpha_1\ne0$, hence instability, with the exception of close to the critical point where $\alpha_1\approx0$.

 \section{Fluctuation-induced renormalization of $\alpha$} \label{sec:Renormalization}



 The previous analysis did not incorporate noise into the picture, we will do so now and demonstrate how, when the noise is included, the renormalized coefficient $\alpha$ will naturally depend on the density $\rho$ as well. Here, we aim mainly to connect our hydrodynamic EOM derived from our Chapman-Enskog expansion (\ref{sec:TonerTuRecovery}) to the results of the linear stability analysis, and so will not perform a full renormalisation procedure on the EOM. Specifically, we will show how the $\beta$ term, together with the noise term with strength $\sigma$, renormalise $\alpha$ around the critical region when $\alpha \approx 0$.

 Firstly, we must understand the macroscopic effect of the noise. Recall that 
 \begin{equation}
 	\sum_{i=0}^{7} \eta_{i} = 0 \ ,
 \end{equation}
 hence there is no contribution to the continuity equation.
 Let's define
 \begin{equation}
 	\bh = \sum_{i=0}^{7} \eta_{i} c \be_{i} \ .
 \end{equation}
 From the definition of $\eta_{i}$ it follows that
 \begin{eqnarray}
 	\left< h_{x} \right> = \left< h_{y} \right> = \left< h_{x} h_{y} \right> = 0 \ , \\
 \fl \qquad	\left< h_{x}(\bx,t) h_{x}(\bx',t') \right> = \left< h_{y}(\bx,t) h_{y}(\bx',t') \right> = c^2 \sigma^2 \delta(\bx-\bx')\delta(t-t') \ .
 \end{eqnarray}
 Our derived \eq \eqref{eqn:DerivedTonerTuMomentum} now becomes
 \begin{equation}
 	\rho \left(	\partial_{t} \bv   +  \bv \cdot \nabla   \bv \right)	
 	= - \nabla P   + \nabla \cdot \rho \nu \left( \nabla \bv + (\nabla \bv)^{T} \right) + \bF_{\rm CIL} + \bh	\ .
 \end{equation}
 
 The calculation mirrors almost exactly the corresponding calculation in the study of incompressible active polar fluids at criticality \cite{Chen2015}, except that the projection operator is absent here since our system is not incompressible. Specifically, the shift in $\alpha$ is given by
 \begin{eqnarray}
 	\delta \alpha &= -\beta \int \frac{\dd^{\rm D} {\bf q}}{(2\pi)^{\rm D}}\frac{\dd \Omega}{2\pi} \frac{\left< |\boldf|^2 \right>}{(\mu q^2+\alpha)^2 +\Omega^2}
 	\\
 	&= -\beta \frac{c^2 \sigma^2}{\rho^2} \int \frac{\dd^{\rm D} {\bf q}}{(2\pi)^{\rm D}}\frac{1}{\mu q^2+\alpha}
 	\\
 	&= -\frac{c^2 \sigma^2 \beta}{\rho^2 \left(\mu \Lambda^2 +\alpha\right)} \frac{S_{\rm D} \dd \ell}{(2\pi)^{\rm D}} \ ,
 \end{eqnarray}
 where  $\boldf = \frac{\bh}{\rho}$, $S_{\rm D} \equiv 2\pi^{{\rm D}/2}/\Gamma ({\rm D}/2)$ is the surface area of a unit sphere in D dimensions, $\Lambda^{-1}$ corresponds to the small length scale cutoff, and $\Lambda\dd \ell$ corresponds to the width of the momentum shell being integrated over in the above coarse-graining process.
 
 Adding this shift to $\alpha$ modifies it to, to $\cO(\alpha)$:
 \begin{equation}
 	\alpha_R = \alpha \left(1- \frac{\sigma }{\rho^2 U_0^2\mu \Lambda^2} \frac{S_{\rm D} \dd \ell}{(2\pi)^{\rm D}}\right) = \alpha\left( 1- C_1 \frac{1}{\rho^2 U_0^2} \right)  \ ,
 \end{equation}
 where we have used the expression for $\beta$ in \ref{eqn:beta}.
 
 The above expression indicates that the critical transition occurs when $U_0$ and $\rho$ are fine tuned such that $\alpha_R =0$ and $\pp \alpha_R /\pp \rho=0$, which can be achieved by fine tuning  $\bar{\rho}$ and $A$ in the expression of $U_0$.  These are given by
 \begin{equation}
 	\bar{\rho} = \frac{3\sqrt{C_1}}{2B}  \sep   A = \frac{4 B^3}{27 C_1}\ .
 \end{equation}
 In our simulations we found $\bar{\rho}_c \approx 0.18$, $A_c \approx 2.75$. These values give $B \approx 0.27, C_1 \approx 0.001$, which is fairly consistent with our actual value of $B = 0.3$.
 In fact we can go further. In \fig 3(a) (and \fig S3) the blue dashed line depicts the full contour for $\alpha_{R}=0$ using these values.  For parameter values below this curve $\alpha>0$ and we have collective motion, with $\alpha<0$ for parameter values above.
 By design this line cuts through our critical point (star), however it also bisects the banding and reverse banding regions, in perfect agreement with our stability analysis; in the vicinity of this curve $\alpha_{0}\approx 0$ and $\alpha_{1} \ne 0$ except at the critical point.

\section{Critical phenomena in active systems}
 We are aware of critical phenomena analysed, either analytically or computationally, in the following active matter systems: 1) self-propelled particles with long-ranged metric-free alignment interactions \cite{Ginelli2010,Peshkov2012}, 2) 
 active L\'{e}vy matter \cite{Cairoli2019active,Cairoli2019hydrodynamics}, 3) non-equilibrium diffusive particles with long-ranged dynamics \cite{Grossmann2016},
  4) incompressible active fluids \cite{Chen2015}, 5) critical motility-induced phase separation \cite{Partridge2019,Siebert2017,Caballero2018}, and 6) self-propelled particles with velocity reversals and alignment interactions \cite{Grossmann2016_PRE94,Mahault2018}. 
Our critical system is different from system 1) to 3) because ours does not have long-range interactions or dynamics, from 4) because our velocity field is not divergence free,
 from 5) because critical MIPS happens in the disordered phase while criticality  in our system occurs at the onset of the ordered phase, from 6) because our system exhibits collective motion while system 6) exhibits only orientational order, but no collective motion. Therefore, we believe that the critical behaviour uncovered here is likely to be novel.

\section{Conclusion \& Outlook}
Through symmetry considerations,
	 we have generalized LBM to simulate dry active fluid systems and uncovered novel phase transitions. 
Looking ahead, we believe that the drastic improvement on computational efficiency will help solve interesting open questions in dry active fluids, such as the dynamic scaling in the incompressible limit \cite{Chen2016}, and universal behaviour in chiral active fluids \cite{Nguyen2014,Denk2016,Liebchen2017}. Furthermore, the fact that LBM are adaptable to complex geometries renders them suitable for investigating interfacial instabilities in a multiphase active fluid system, such as in the context of tissue regeneration \cite{Basan2013,Zimmermann2014,Nesbitt2017}. Finally, we believe our method can be straightforwardly extended to higher dimensions. 


\ack
D.N. was supported by the EPSRC Centre for Doctoral Training in Fluid Dynamics Across Scales (grant
EP/L016230/1). 

\appendix

\section{Simulation Algorithm} \label{sec:Algorithm}
\subsection{Conventional Lattice Boltzmann Methods} \label{sec:LBM}
Lattice Boltzmann methods ({\bf LBM}) refers to solving the Boltzmann equation on a lattice. The Boltzmann equation itself concerns the temporal evolution of a  distribution function  $f(t,\bbr,\bxi)$, which corresponds to  the mass density of matter moving with velocity $\bxi$ at location $\bbr$ and time $t$. 

The typical variables of interest  are
	\begin{eqnarray}
		 \rho(t, \bbr) &=& \int  \dd^{\rm D}\bxi \  f(t,\bbr,\bxi)  \ , \label{eqn:macroscopic_rho_cts} \\
		\rho(t, \bbr) \bu(t, \bbr) &=& \int\dd^{\rm D}\bxi \    f(t,\bbr,\bxi)\bxi
		\ , \label{eqn:macroscopic_u_cts}
	\end{eqnarray}
where D is the spatial dimension, $\rho$ is the mass density field, and $\bu$ is the velocity field.

Discretising $\bxi$  into a finite number of velocity vectors, directed towards neighbouring spatial nodes, naturally constructs a lattice in the domain. There are many suitable choices of lattice, but they must contain  enough degrees of freedom to preserve  the moments of the system in question according to the desired level of accuracy \cite{Frisch1986,Chen2008}. The DQ notation is conventional for describing the lattice structure of the domain, where D is again the dimension, and Q is the number of nodes each is connected to, including itself. Typically  a `zero' vector is included in the discrete directional space to help balance higher moments.
Here, we will focus on a D2Q7 lattice, which is a triangular lattice in two dimensions with a discretised distribution function $f_{i}(t,\bbr)$, where $i = 0,1,...,6$. The corresponding vectors are ${\be}_{0} = \bzed$, and  ${{\be}_{i}} = \cos [(i-1)\pi/6] \hat{\bx} +\sin [(i-1)\pi/6] \hat{{\bf y}}$ for $i>0$.  The local mass density and velocity are thus given by
\begin{equation}
	\rho = \sum_{i=0}^{6} f_{i} 
	\sep
	\bu = \frac{1}{\rho}\sum_{i=0}^{6} f_{i} c \be_{i}
	\ , 
\end{equation}
where $c$ is the ratio of the lattice spacing $\Delta x$ to the time increment $\Delta t$.
For accurate solutions velocities must obey $|\bu| \ll  c$, which is known as the low Mach number assumption.
Typically scales in the LBM scheme are chosen such that $c=1$.

The discretized Boltzmann equation can then be evolved 
using the  Bhatnagar-Gross-Krook collision operator \cite{Bhatnagar1954}
\begin{equation} \label{eqn:fevolution}
	f_{i}( t + \Delta t , \bbr + c \be_{i} \Delta t ) - f_{i}(t,\bbr )  =  
	- \frac{1}{\tau} \left[ f_{i}(t,\bbr )   -  f_{i}^{eq}(t,\bbr ) \right]  
	\ ,
\end{equation}
where $f_{i}^{eq}(t, \bbr )$ is a lattice dependent steady-state (equilibrium) distribution and $\tau$ is a relaxation parameter. When simulating the Navier Stokes equations, $\tau$ is linearly proportional to the viscosity. The evolution of this equation is typically split into two processes--the streaming step and the collision step, the latter of which is effectively a local redistribution.
The design of the steady-state distribution is highly flexible and the only requirement is that it respects the symmetries and conservation laws imposed on the system. 
For the Navier-Stokes equation, $f_{i}^{eq}$ is obtained by discretising the Boltzmann distribution for velocities at thermal equilibrium \cite{He1997_PRE} and in a D2Q7 framework is given by
\begin{equation}
	f_{i}^{eq} = w_i \rho \left[ 1	 +  4\frac{\be_i \cdot \bu}{c} + 8\frac{(\be_i \cdot \bu)^2}{c^2}- 2\frac{|\bu|^2}{c^2}	  \right] 
	\ , \label{eqn:feqD2Q7}
\end{equation}
where $w_i$ are lattice specific weights: $w_0 = 1/2$ and $w_{i\ne0} = 1/12$.  This discretisation is possible through the low Mach number assumption $|\bu| \ll c$.

\subsection{Summary of modified lattice Boltzmann model algorithm}
A single time step is partitioned into the following steps
\begin{enumerate}
	\item
	Streaming: $f_{i}^{*}(t + \Delta t, \bbr +  c \be_{i} \Delta t) = f_{i}(t,\bbr)$.
	\item
	Calculate $\rho$, $\bu$ from $f_{i}^{*}$.
	\item
	Calculate $\bu^{*}$ and hence $f_{i}^{\rm SS}$.
	\item
	Redistribution (collision):
	$f_{i}  =  f_{i}^{*}  - \frac{1}{\tau} \left[ f_{i}^{*}-  f_{i}^{\rm SS} \right]$.
	\item
	Add noise:
	$f_{i} \mapsto f_{i} + \eta_{i}$.
	\item 
	Correct any negative values of $f_{i}$ (see Section \ref{sec:NegativeCorrection}).
\end{enumerate}

\subsection{Negative Correction} \label{sec:NegativeCorrection}

In our modified LBM, we have included stochastic noise as detailed in the main text. It is possible for one or more of the $f_{i}$ values to become negative in this process. This is not permitted, hence we correct this using the following procedure at each lattice site.

\begin{enumerate}
	
	\item
	If $f_{0}<0$, simply set $f_{0} =0$.
	
	\item
	If $f_{i}<0$ for $i>0$, then take $j$, such that $\be_{j}$ is the reverse direction of $\be_{i}$ ( $1\longleftrightarrow 4, 2\longleftrightarrow5,3\longleftrightarrow6$), and add $|f_{i}|$ to $f_{j}$, then set $f_{i}=0$.
	
	\item
	By now $f_{i}\ge0$ for all $i$, however it is possible that the density has changed. This is corrected by rescaling
	\begin{equation}
		f_{i} \mapsto \frac{\rho}{\sum_{j} f_{j}} f_{i} \ .
	\end{equation}			
\end{enumerate}

Also note that due to both the noise and the rescaling of $\bu$ in $f_{i}^{\rm SS}$, it is possible that 
\begin{equation}
	f_{i}(t,\bbr)  - \frac{1}{\tau} \left[ f_{i}(t,\bbr) -  f_{i}^{\rm SS}(t,\bbr) \right] < 0
	\ ,
\end{equation}
when $\tau < 1$, even if $f_{i}>0$ for all $i$. Therefore to ensure that no values become negative, we only use $\tau \ge 1$. Different values of $\tau$ were tested, yielding results that were qualitatively similar to those described in the letter.

\section{Lattice Boltzmann to Toner-Tu equations via Chapman-Enskog expansion} \label{sec:TonerTuRecovery}
In this section we will demonstrate that our LBM simulates a version of the Toner-Tu equations. By expressing our active LBM in the form of a conventional LBM with a body force term, we can use the results of \cite{Li2016} to derive the macroscopic equations.

Let's introduce the notation
\begin{equation}
	\xi_{i}(\rho,\bu) = w_i \rho \left[ 1	 +  4\frac{\be_i \cdot \bu}{c} + 8\frac{(\be_i \cdot \bu)^2}{c^2}- 2\frac{|\bu|^2}{c^2}	  \right] 
	\ .
\end{equation}
Hence in the conventional LBM $f_{i}^{eq} = \xi_{i}(\rho,\bu)$ and in our active model $f_{i}^{\rm SS} = \xi_{i}(\rho,\bu^{*})$.

The evolution equation becomes
\begin{equation}
	f_{i}( t + \Delta t , \bbr +  \be_{i} \Delta t ) - f_{i}(t,\bbr )  
	=   - \frac{1}{\tau} \left[ f_{i}   -  \xi_{i}(\rho,\bu) \right]  + F_{i} 
	\ , \label{eqn:evolutionWithF}
\end{equation}
where
\begin{equation}
	F_{i} = \frac{1}{\tau} \left[ \xi_{i}(\rho,\bu^{*})   -  \xi_{i}(\rho,\bu) \right] \ .
\end{equation}
Note that
\begin{eqnarray}
	\sum_{i=0} F_{i} = 0 \sep \sum_{i=0} F_{i} c \be_{i} = \rho \frac{\bu^{*} - \bu}{\tau} = \Delta t \  \bF
	\ .
\end{eqnarray}
As in \cite{Li2016}, we define the `actual velocity' according to
\begin{equation} \label{eqn:actualVelocity}
	\bv = \bu + \frac{ \Delta t \bF}{2\rho} = \bu - \frac{1}{2\tau} \left(\bu - \bu^{*} \right)
	\ .
\end{equation}
After expressing \eq \ref{eqn:evolutionWithF} in terms of the actual velocity,
\begin{equation}
	f_{i}( t + \Delta t , \bbr +  \be_{i} \Delta t ) - f_{i}(t,\bbr )  
	=  
	- \frac{1}{\tau} \left[ f_{i}   -  \xi_{i}(\rho,\bv) \right] + \tilde{F}_{i} 
	\ , \label{eqn:evolutionWithHatF}
\end{equation}
where 
\begin{equation} \label{eqn:FTilde}
	\tilde{F}_{i}  =  F_{i} +  \frac{1}{\tau} \left[ \xi_{i}(\rho,\bu) - \xi_{i}(\rho,\bv) \right] = \frac{1}{\tau} \left[ \xi_{i}(\rho,\bu^{*}) - \xi_{i}(\rho,\bv) \right]  \ ,
\end{equation}
we can now use the results of \cite{Li2016} to determine our macroscopic equations.
They are the continuity equation
\begin{equation}
	\partial_{t} \rho  + \nabla \cdot \left( \rho \bv \right) = 0 \ ,
\end{equation}
and the momentum equation
\begin{equation} 
	\partial_{t} \left( \rho \bv \right)  +  \nabla \cdot  \left( \rho \bv \bv \right)  =  - \nabla P  + \nabla \cdot \tilde{\Pi} + \bF + \nabla \cdot \Theta \ ,
\end{equation}
where
\begin{equation}
	\tilde{\Pi} = \rho \nu \left( \nabla \bv + (\nabla \bv)^{T} \right)
	\ ,
\end{equation}
and
\begin{equation}
	\Theta = \left( \tau - \frac{1}{2} \right) \Delta t \left( \bv \bF + \bF \bv \right)  - \tau \sum_{i} \be_{i} \be_{i} \tilde{F}_{i} \ .
\end{equation}
Several $O(\bu^3)$ terms have been ignored, consistent with a small Mach number assumption.
The pressure $P=c_{s}^2 \rho$ and the viscosity
\begin{equation}
	\nu = \frac{2\tau - 1}{2}c_{s}^{2}\Delta t \ ,
\end{equation}
both feature the effective speed of sound $c_{s}$ which depends on the choice of lattice. In the case of a D2Q7 lattice, $c_{s} = c/2$.
From \eq \eqref{eqn:FTilde},
\begin{equation}
	\tau \sum_{i} \be_{i} \be_{i} \tilde{F}_{i} = \rho \bu^{*} \bu^{*} - \rho \bv \bv \ .
\end{equation}

Rearranging \eq \eqref{eqn:actualVelocity}, we can express both $\bu$ and $\bu ^{*}$ in terms of $\bv$ by
\begin{equation}
	\bu = \frac{1}{2\tau - 1} \left( 2 \tau \bv  -\bu^{*} \right)
	\ .
\end{equation}
Thus 
\begin{equation}
	\bF = \frac{\rho}{\tau \Delta t} \left( \bu^{*} - \bu \right)  = \frac{\rho}{\Delta t}  \frac{2  }{2\tau - 1} \left( \bu^{*} - \bv \right) \ .
\end{equation}
Substituting all of these into $\Theta$, and factorizing gives
\begin{equation}
	\Theta = - \rho \left( \bu^{*} - \bv \right) \left( \bu^{*}- \bv \right).
\end{equation}
Clearly $\nabla \cdot \Theta$ has parallels with the term $\nabla \cdot (\rho\bv \bv)$. However, for our simulation to behave smoothly, we assume that $(\bu^{*} - \bv) \sim O(\varepsilon)$ where $\varepsilon$ is small, thus we can safely ignore $\nabla \cdot \Theta$, as it is of order $ O(\varepsilon^2)$.
Therefore the resulting momentum equation is
\begin{equation}	\label{eqn:DerivedTonerTuMomentum}	
	\rho \left(	\partial_{t} \bv   +  \bv \cdot \nabla   \bv \right)	
	= - \nabla P   + \nabla \cdot \rho \nu \left( \nabla \bv + (\nabla \bv)^{T} \right) + \bF 	.
\end{equation}
This is valid for both the MIPS model and the CIL model; the differences stem from the choice of $\bu^{*}$ in $\bF$. Both are versions of the Toner-Tu equations \cite{Toner1995,Toner1998,Toner2005,Toner2012}.

\subsection{MIPS model}
For our MIPS model we use $\bu^{*}$ as given in \eq \eqref{eqn:bu_mips}, thus
\begin{equation} \label{eqn:Fmips}
	\bF_{\rm MIPS} =  \frac{2 \rho }{\Delta t (2\tau - 1)} \left( C \left(1 - \frac{\rho}{\rho^{*}} \right)\nabla \rho  - \bv \right) \ .
\end{equation}
Since $\nabla P = c_{s}^2 \nabla \rho$, the first term on the RHS can be viewed as an effective pressure in which the coefficient varies with density. Crucially, for $\rho<\rho^{*}$ this coefficient is positive, hence the effective pressure acts in the opposite manner from the conventional pressure, driving material towards areas of high density instead of away from them. 

The second term can be expressed as $\alpha \rho \bv$ with
\begin{equation}
	\alpha = -\frac{2}{2\tau -1} < 0 \ .
\end{equation}
As $\alpha$ is negative, this term acts like friction, reducing the velocity in the system. As such $\bv$ decays to a value determined by a balance between the effective pressure, the conventional pressure and the stochastic noise. This balance is also seen in \eq \eqref{eqn:MIPS_Macro_J}, thus our method is consistent with other models of MIPS.

\subsection{CIL model}
For our CIL model we use $\bu^{*}$ as given in \eq \eqref{eqn:bu_CIL}. Note that
\begin{equation}
	\bu^{*} = U_0 \frac{\bu }{|\bu|} = U_0 \frac{\bv}{|\bv|} 
	\ .
\end{equation}
Thus
\begin{eqnarray}
	\bF_{\rm CIL} =  \frac{2 \rho }{\Delta t (2\tau - 1)} \left(  U_0 \frac{\bv}{|\bv|} - \bv \right) 
	\ .
\end{eqnarray}
$\bF_{\rm CIL}$ corresponds to a linearised version of 
$\rho\left[\alpha(\rho)- \beta |\bu|^2 \right]\bu$ in the Toner-Tu equations,
where the linearisation has occurred about the stationary points $|\bu| = \sqrt{\alpha/\beta}$.  Matching the terms gives
\begin{eqnarray} \label{eqn:beta}
	\alpha = \frac{1}{\Delta t \left( 2 \tau - 1 \right)}  \sep  \beta = \frac{\alpha}{U_0^2} \ .
\end{eqnarray}
These are not well defined when $U_0 = 0$. In this case 
\begin{eqnarray}
	\frac{\bF_{\rm CIL}}{\rho} = - \frac{2  }{\Delta t (2\tau - 1)}  \bv 
	\ .
\end{eqnarray}
From this we can deduce nothing about $\beta$, but conclude that
\begin{eqnarray}
	\alpha =- \frac{2}{\Delta t \left( 2 \tau - 1 \right)}  \ ,
\end{eqnarray}
as in the MIPS model.
These values of $\alpha$ motivate the idea that $\alpha \rho \bu$ is a driving force in the system; $\alpha > 0$ gives a non-zero flow $U_0$, whereas if $\alpha < 0$ flow will decay to $\bu =\bzed$.

\section{Simulation procedure for results} \label{sec:SimulationDetails}

\subsection{MIPS Model}

We are primarily concerned with the effects of varying the mean density and noise strength; we held the remaining parameters constant with the values in Table \ref{tab:ParametersMIPS}. 
Note that our values are non-dimensional.
\begin{table}[h]
	\begin{center}
		\begin{tabular}{c c} 
			Parameter		&	Value	\\
			\hline
			$\tau$			&	2 		\\
			$C$				&	50		\\
			$\rho_{*}$		&	0.2		\\
			$\Delta x$		& 	1		\\
			$U_{\rm max}$	&	0.3		\\
			$c$				&	1 		\\
			\hline
		\end{tabular}
	\end{center}
	\caption{Parameters used to obtain the results depicted in \fig \ref{fig:MIPS}. Values are non-dimensional.}
	\label{tab:ParametersMIPS}
\end{table}
To obtain the still frames in \fig \ref{fig:MIPS}(a)-(d), we used a lattice of $180\times60$ nodes, and the parameters:
\begin{center}
	\begin{tabular}{c c c} 
		&	$\bar{\rho}$	& $\sigma$	\\
		\hline
		(a)		&	0.10 	&	0.031	\\
		(b)		&	0.10	&	0.014 	\\
		(c)		&	0.18	&	0.014	\\
		(d)		&	0.10	&	0.014	\\
		\hline
	\end{tabular}
\end{center}

To generate a phase diagram we reduced the system size to $90\times 30$ nodes, due to the long runtimes required to achieve the state in \fig \ref{fig:MIPS}(d). Doing so had no obvious effect on the qualitative behaviour of the system, hence we concluded that this system size is large enough to account for any finite-size effects.
We fixed $\bar{\rho}=0.12$, and slowly varied $\sigma$.
The simulation was initialized with $\sigma = 1/70\approx 0.014$, in a state close to the theoretical equilibrium. It was allowed to evolve for a further $10^6$ time steps before measurements were taken. A sample was then taken every $10^4$ time steps, for $100$ samples. After taking the final sample, the noise strength was increased by $\delta \sigma = 1/700$. The system was allowed to relax for another $10^6$ time steps before the measurements resumed at the same rate as before. After a further $100$ samples the noise strength was increased again by $\delta \sigma$ and this process was repeated until $\sigma = 3/70$.

To systematically determine whether or not phase separation had occurred, we quantified the large scale density variation by a parameter $\denseVar$ (see \ref{sec:RegimeCategorisation}). \fig \ref{fig:MIPS}(e) shows the mean values of $\denseVar$ for the $100$ samples for each value of $\sigma$. The error bars are the estimated standard deviation of the samples (not the sample mean), indicating the spread of the sample distribution. 
Large values of $\denseVar$ indicate phase separation, but given that the system is noisy, $\denseVar$ is never exactly zero, hence we must impose a cutoff. The value of $\denseVar$ falls with increasing noise. Furthermore it falls faster and faster, until it quickly levels out around $\sigma = 0.0275$, at which point the standard deviation of the samples drops as well. From inspection of the samples, this is the point at which the phase separation ceases to be obvious. Hence we use $\denseVar_{ \rm threshold} = 0.0430$, with $\denseVar>\denseVar_{ \rm threshold}$ taken to be phase separated.

Using this condition to identify cases in which phase separation had occurred, we measured the density in the liquid and gas phases. We made the assumption that the system had reached the theoretical steady state, so that the dense liquid phase is entirely contained in a single vertical strip. To obtain the typical density value, both within and outside of the strip, we averaged the density profile over a rectangle spanning the entire $y$-axis, but only $20\Delta x$ wide. This rectangle was shifted along the entire domain, with the highest and lowest values obtained taken as the density of the liquid and gas phases respectively.


\subsection{CIL Model: General notes}

We used a $180\times60$ node triangular lattice with periodic boundary conditions. Increasing the size to $360\times120$ nodes produced no obvious qualitative change in behaviour, hence we deduced that $180\times60$ was large enough to negate finite-size effects. Every simulation was initialized by choosing 2 uniformly distributed random numbers for each lattice site---the first deciding the initial direction for $\bu$ with each direction in $[0,2\pi]$ equally likely, and the other choosing the initial density such that it falls within $\pm10$\% of $\bar{\rho}$, the prescribed average density. The initial magnitude of $\bu$ is determined by \eq \eqref{eqn:U0}, 
and the initial distribution given by $f_{i}^{\rm SS}$ in \eq \eqref{eqn:fssD2Q7}. The system was then evolved 10000 steps before a ``run" would formally begin. This marks the end of the initialization process. 

As detailed below, during the course of each run the mean density would be deliberately changed by a small amount $\delta \bar{\rho}$. On each occurrence, this was achieved by adding $\delta \bar{\rho} / 7$ to every $f_{i}$ across the lattice---the lattice was not reinitialized. For measurements involving averaging over multiple runs the initialization process was repeated between each run.

With the exception of $A$ and $\bar{\rho}$, the parameters were held constant with $\Delta x =1$, $\Delta t =1$, $c=1$, $B = 0.3$,  $\tau = 1$, and  $\sigma = 1/70$.  As with the MIPS model, these parameters are all non-dimensional. In conventional LBM $\tau$ is linearly proportional to the viscosity of the system. We expect a similar relationship here, but the effect of viscosity is not the main focus of this study.

\subsection{Simulation procedure in Fig. \ref{fig:Hysteresis}}

For \fig \ref{fig:Hysteresis}(a) $A=2$, and only $\bar{\rho}$ is varied. The goal was to detect any hysteresis effect when adiabatically increasing or decreasing the density.
For the increasing density run, 
we start with  $\bar{\rho} = 0.09$ and increase it by 
$|\delta \bar{\rho}| = 0.0025$ every $t_{\rm sample}$ steps. A sample was taken just before each density change. For higher densities the system would reach equilibrium faster, hence in order to be able to detect the hysteresis effect between HO-RB and RB-HD, a lower value of $t_{\rm sample} = 500$ was used  for $\bar{\rho}>0.2$, with  $t_{\rm sample} = 1500$ otherwise. 
This is because in a finite system, running the system long enough would eliminate all hysteresis effects. 

For the decreasing density run, 
we start with  $\bar{\rho} = 0.35$ and again decrease it by  $|\delta \bar{\rho}| = 0.0025$ every $t_{\rm sample}$ steps, which is again 500  for $\bar{\rho}>0.2$ and  1500 otherwise. The system was reinitialized between increasing and decreasing runs. As a result, they should be treated as independent.

The whole increasing and decreasing density cycle was repeated over  50 runs. The order parameter $\varphi$ was evaluated for each sample, and the values for samples at equal densities $\bar{\rho}$ taken in the same direction of the cycle
were collated and averaged. The error bars are the standard deviation of $\bar{\varphi}$, that is, the estimate of the standard deviation of the samples $\sigma_{\varphi}$ scaled by the square root of the sample size,  $\sigma_{\varphi}/\sqrt{50}$.

\subsection{Simulation procedure in Fig. \ref{fig:PhaseDiagram}}

For \fig \ref{fig:PhaseDiagram} $A$ was varied as well as $\bar{\rho}$. Results were obtained by fixing $A$ then slowly varying $\bar{\rho}$, much as for \fig \ref{fig:Hysteresis}(a) except: there were no descending runs, $\Delta \rho = 0.01$,  and  $ t_{\rm sample} = 10000$, which is typically sufficient for the system to reach its steady state.
Each run was repeated 10 times  resulting in 10 independent samples for each point $(\bar{\rho},A)$.  For each of these samples, the parameters $\varphi, \denseVar, \zeta$
(see Section \ref{sec:RegimeCategorisation}) were calculated, then averaged for each point $(\bar{\rho},A)$. The resulting values were used to deduce the behaviour of the system for those parameters.

\section{Regime Categorisation in the phase diagram} \label{sec:RegimeCategorisation}

To systematically categorise the system's behaviour at each of our data points, we evaluate several parameters.

The order parameter $\varphi$ measures the global velocity alignment.
Since we have a finite system, even in the most disordered phase $\varphi$ is never exactly zero, and as it is strictly positive it cannot average to zero either. Therefore we had to impose a cutoff to distinguish between a completely disordered state and one with an ordered phase (including the B and RB phases).
For our CIL model we deduced the cutoff to be $\varphi_{\rm threshold} = 0.024$, with  $\varphi <\varphi_{\rm threshold}$ marked as HD.
The MIPS model has no mechanism for collective motion, thus $\varphi \approx 0$ for all parameter values.

To determine whether or not phase separation had occurred, we quantified the
large scale density variation by a parameter $\denseVar$. In both the MIPS and CIL models, due to the aspect ratio of the geometry and the periodic boundary conditions, the steady state of the phase-separated regimes manifested as strips of high density, parallel to the $y$-axis, as demonstrated in \fig \ref{fig:MIPS}(d) and \fig \ref{fig:Hysteresis}(d). In these cases, the system is largely invariant in the $y$-direction. 
Averaging over the density in that direction,
\begin{equation}
	\rho_y(x) = \frac{1}{L_y} \int_{0}^{L_y} \dd y \ \rho(x,y) \ ,
\end{equation}
thereby reduces the effects of the small scale stochastic fluctuations. We then took $\denseVar$ to be the standard deviation of $\rho_y$.

As with $\varphi$ we had to impose a cutoff to distinguish between phase separated regions and local noise. For our CIL model, we chose this to be $\denseVar_{ \rm threshold} = 0.012$, with $\denseVar > \denseVar_{ \rm threshold}$ considered to be either B or RB.

The background colour in the phase diagram in \fig \ref{fig:PhaseDiagram}(b) corresponds to the value of $\denseVar$, and demonstrates that our chosen cutoff value is appropriate. The two strips of yellow (high $\denseVar$) clearly correspond to the B and RB regimes. These strips narrow as they approach each other, and vanish entirely at the critical point (star).


The two aforementioned quantities, $\varphi$ and $\denseVar$,  do not enable us to distinguish between the B and RB phases. Therefore we also measured the following parameter:
\begin{equation}
	\zeta = \frac{\sum_{ \bbr \in {\rm lattice}} \rho(\bbr)[B - |\bu(\bbr)|] }{\sum_{\bbr \in {\rm lattice}} \rho(\bbr)} 
	\ ,
\end{equation}
where $\zeta$
is large if high density regions are stationary, as in RB. Hence we identified RB values with 
$\zeta > \zeta_{\rm threshold} = 0.1$.
In summary, for the CIL model, data points were classified using the following rules
\begin{center}
	\begin{tabular}{c c c c} 
		Regime	& 	$\varphi$ 	&	$\denseVar$		&	$\zeta$		\\
		\hline
		HD		&	$< \cutoffCILAlign $	&		--			&	--			\\
		HO		&	$\ge \cutoffCILAlign$	&		$\le \cutoffCILDense$		&	--			\\
		B		&	$\ge \cutoffCILAlign$	&		$> \cutoffCILDense$		&	$\le \cutoffCILZeta$	\\
		RB		&	$\ge \cutoffCILAlign$	&		$> \cutoffCILDense$		& 	$>\cutoffCILZeta$		\\
		\hline
	\end{tabular}
\end{center}	

Visual inspection of the samples at several distinct points in the phase diagram shows the categorization to be largely accurate.

For the phase diagrams in \fig \ref{fig:PhaseDiagram}, anything above the red line $B -A \bar{\rho}^2 = 0$ was marked as motionless.

\section{Representation of static configurations} \label{sec:VisualizationScheme}

Each of the plots in \fig \ref{fig:Hysteresis}(d)  and \fig \ref{fig:MIPS}(a)-(d) are scatter plots of $180\times60$ circles, arranged in the triangular lattice configuration, as is \fig \ref{fig:PhaseDiagram}(c) but for a lattice of $720\times240$ nodes.  They represent the density and velocity fields  throughout our lattice for a single time step.
The colour of each dot corresponds to the direction of the macroscopic velocity $\bu$ measured at that site, according to the  inset colour wheel. 
As our system is two dimensional a single angle (colour) is sufficient to determine the direction.
The area of each dot is assigned by the following formulae
\begin{equation}
	a = a_{0} \left(\frac{\rho-\rho_{\min}}{\rho_{\max}-\rho_{\min}}\right)^{3/2} + \epsilon \ ,
\end{equation}
where $\rho_{\max},(\rho_{\min})$ is the maximum (minimum) value of $\rho$ over all of the lattice sites in that frame, $a_0$ is an appropriately chosen value for the size of the biggest circle, which depends on the spacing of the particular plot. $\epsilon$ is a small value to ensure no values are exactly zero. 

In \fig \ref{fig:Hysteresis}(d) the samples are obtained from one of the increasing runs used to create \fig \ref{fig:Hysteresis}(a), specifically sampled with the mean densities $\bar{\rho}$:
\begin{center}
	\begin{tabular}{l l} 
		HD	&	0.105 \\	
		B	&	0.1425 \\	
		HO	&	0.185 \\	
		RB	&	0.2875 \\	
	\end{tabular}
\end{center}

\fig \ref{fig:PhaseDiagram}(c) is created with the values $A=2.7$, $\rho =0.175$ and a lattice size of $720\times240$, with all other parameters the same as before. We have chosen a higher resolution to better illustrate the critical behaviour. This particular frame is $100000$ steps after initialization.

\section{Movies} \label{sec:Movies}
Four movies are provided 
to demonstrate the behaviour of the system as shown in \fig \ref{fig:Hysteresis}(d) and \fig \ref{fig:PhaseDiagram}(c), using the same visual representation method. 
The movies are available at https://www.dropbox.com/sh/0q16xf2t9tfi8h1/AACC8AYrVLwhkYegVimgBdega?dl=0.

For each of the movies the system was initialized with $A=2$ and $\bar{\rho}$ again given by
\begin{center}
	\begin{tabular}{l l l} 
		M1	& HD	&	0.105 \\	
		M2	& B	    &	0.1425 \\	
		M3	& HO	&	0.185 \\	
		M4	& RB	&	0.2875 \\	
	\end{tabular}
\end{center}
respectively.
A frame was captured every 5 time steps. Each movie is constructed of 200 frames displayed at 12 frames per second.


\section*{References}

\bibliographystyle{unsrt}


\end{document}